\documentclass[11pt]{article}
\usepackage{subfig}
\usepackage{epsfig}
\usepackage{times}
\usepackage{amsfonts}
\usepackage{amsmath}
\usepackage{graphicx}
\usepackage{url}
\usepackage{amssymb}
\usepackage[text={6in,8.5in},centering]{geometry}

\title{Training a Binary Classifier with the \\
Quantum Adiabatic Algorithm}

\author{Hartmut Neven \\
{\it Google, neven@google.com} \\
\\
Vasil S. Denchev\\
{\it Purdue University, vdenchev@purdue.edu} \\
\\
Geordie Rose and William G. Macready\\
{\it D-Wave Systems, {rose,wgm}@dwavesys.com}
}
\begin{document}
\maketitle
\begin{abstract}
\par This paper describes how to make the problem of binary classification amenable to quantum computing. A formulation is employed in which the binary classifier is constructed as a thresholded linear superposition of a set of weak classifiers. The weights in the superposition are optimized in a learning process that strives to minimize the training error as well as the number of weak classifiers used. No efficient solution to this problem is known. To bring it into a format that allows the application of adiabatic quantum computing (AQC), we first show that the bit-precision with which the weights need to be represented only grows logarithmically with the ratio of the number of training examples to the number of weak classifiers. This allows to effectively formulate the training process as a binary optimization problem. Solving it with heuristic solvers such as tabu search, we find that the resulting classifier outperforms a widely used state-of-the-art method, AdaBoost, on a variety of benchmark problems. Moreover, we discovered the interesting fact that bit-constrained learning machines often exhibit lower generalization error rates. Changing the loss function that measures the training error from 0-1 loss to least squares maps the training to quadratic unconstrained binary optimization. This corresponds to the format required by D-Wave's implementation of AQC. Simulations with heuristic solvers again yield results better than those obtained with boosting approaches. Since the resulting quadratic binary program is NP-hard, additional gains can be expected from applying the actual quantum processor.
\end{abstract}
\section{Introduction}
Many problems in machine learning map onto optimization problems that are
formally NP-hard \footnote{From the formal NP-hardness of a class of problems
does not follow that problem instances encountered in practice are
computationally difficult. However, experience tells us that this is usually
the case for the learning problems at hand.}. Consequently, large areas of the field are concerned with
simplifications and relaxations that make the resulting optimization
problems computationally tractable. This has resulted in heuristic tools
that are useful in practice but whose quality is inferior compared to
results obtained by solving the original problem. Moreover, the wealth of
heuristic methods requires that the practitioner needs to select the most suitable approach on a case by case basis. Adiabatic quantum computing is a new method that draws
on quantum mechanical processes that promises to solve hard discrete
optimization problems better than possible with classical algorithms
\cite{Farhi2000}\cite{Farhi2001}. Thus it offers an opportunity to tackle
hard machine learning problems heads on. This paper investigates how this
novel method can be applied to a basic problem in machine learning:
constructing a binary classifier from a dictionary of feature detectors.
\section{Training a binary classifier}
We study a classifier of the form
\begin{equation} y = H(x)={\rm sign}\displaystyle \left(\sum_{i = 1}^{N} w_i h_i(x)\right) \hbox{ , } \end{equation}
where $x \in \mathbb{R}^M$ are the input patterns to be classified, $y \in \{-1,1\}$ is the output
of the classifier, the $h_i:x \mapsto \{-1,1\}$ are so-called weak classifiers or features detectors, and the $w_i \in [0, 1]$ are a set of weights to be optimized.
$H(x)$ is known as a strong classifier.
\par
\begin{figure}[h]
  \begin{center}
    \includegraphics[scale=0.7]{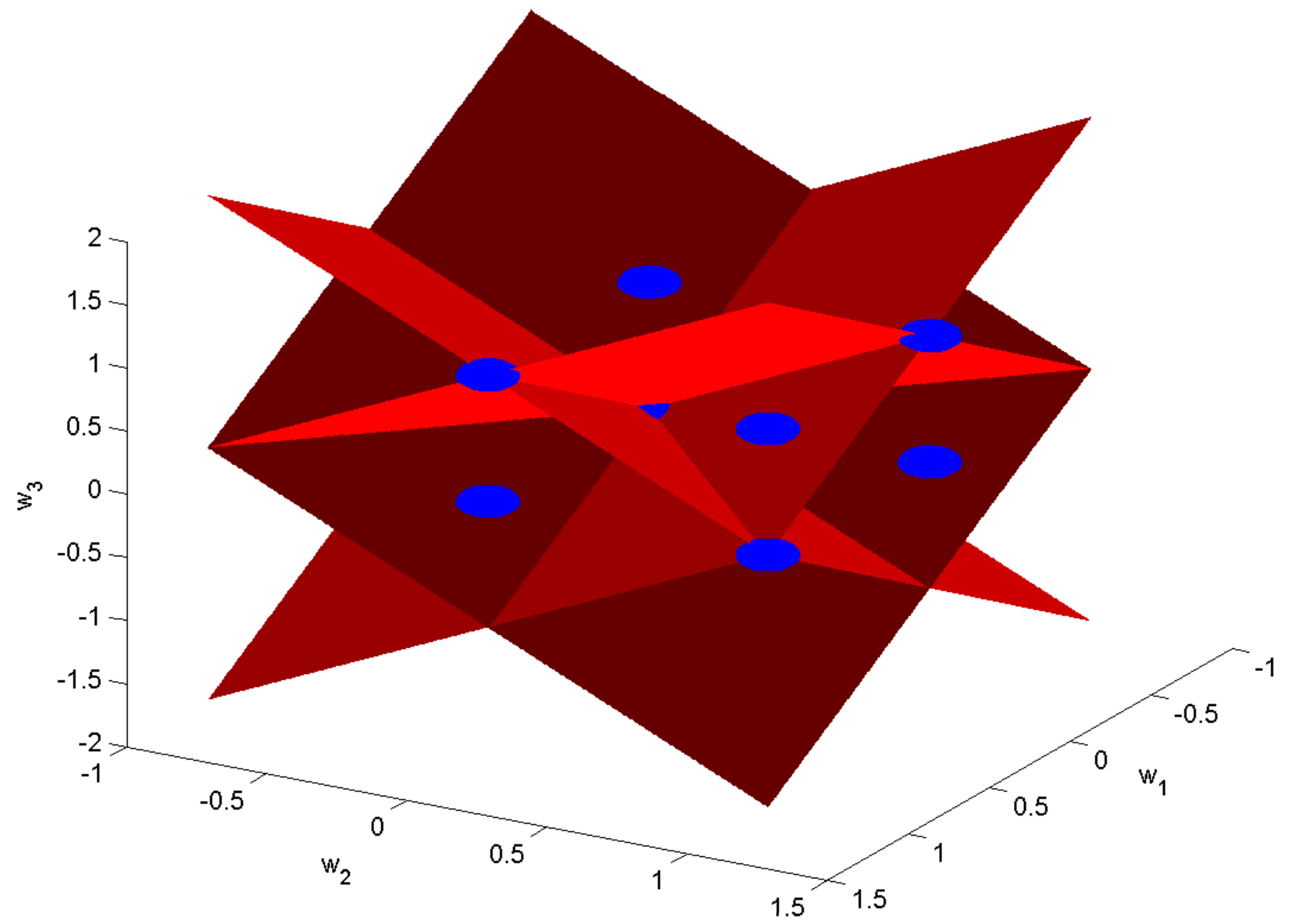}
    \caption{Arrangement of the diagonal hyperplanes that define the solution spaces for selecting $w^{opt}$.
      Depicted is the situation for $N=3$, which yields 14 regions. The number of solution chambers grows
      rapidly -- $N=4$ leads to 104 and $N=5$ to 1882 regions. Here all possible hyperplanes are shown. However in practice $S$ training samples will only invoke a small fraction of the $2^{N-1}$ possible hyperplanes. The blue dots are the vertices of a cube placed in the positive quadrant with one vertex coinciding with the origin. They correspond to weight configurations that can be represented with one bit. Multiple bits would give rise to a cube shaped lattice.}
    \label{fig:3d}
  \end{center}
\end{figure}

 Training, i.e. the process of choosing the weights $w_i$, proceeds by simultaneously minimizing two terms. One term measures the error over a set
of $S$ training examples $\{(x_s, y_s) | s = 1, \ldots, S\}$. A natural choice is 0-1 error, which counts the number of misclassifications over the training set.
\begin{equation} L(w) = \displaystyle \sum_{s = 1}^{S} {\bf H}\left( - y_s \sum_{i = 1}^{N} w_i h_i(x_s)\right) \hbox{ , } \end{equation}
where ${\bf  H}$ is the Heaviside step function. $L(w)$ is referred to as the loss function. The second term is known as regularization, $R(w)$, and it ensures that the classifier does not become too complex. Classifiers with high complexity tend to classify the examples in the training set with low error but do not do well on independent test sets. The phenomenon of a classifier achieving a small training error but yielding a large generalization error is known as overlearning. A simple choice for the regularization term is based on the 0-norm, $\parallel w \parallel _0$, which gives the number of non-zero weights:
\begin{equation} R(w) = \lambda \parallel w \parallel _0 = \lambda \sum_{i = 1}^{N} {w_i}^0\end{equation}
Therefore, training amounts to solving the following minimization problem:
\begin{eqnarray} w^{opt} &=& \arg\min_{w} \left(L(w)+ R(w) \right) \nonumber \\
&=& \arg\min_{w} \left( \sum_{s = 1}^{S} {\bf H}( - y_s \sum_{i = 1}^{N} w_i h_i(x_s)) + \lambda \sum_{i = 1}^{N} {w_i}^0 \right) \hbox{ , }\end{eqnarray}
where $\lambda$ controls the relative importance of the regularization. Due to the non-convexity of the loss function, the resulting optimization problem is suspected to be NP-hard. Even if we were to choose a convex loss function, (4) is likely to remain an NP-hard problem due to the choice of the 0-norm for the regularization term \cite{Zhang}. The choice of the 0-norm is attractive since it explicitly enforces sparsity, i.e. it drives many of the $w_i$ to zero. This is not only associated with good generalization but also fast execution during the performance phase. Each contribution to the overall loss, i.e. the per sample loss, ${\bf H}( - y_s \sum_{i = 1}^{N} w_i h_i(x_s))$, enforces an inequality constraint:
 \begin{equation}
 y_s \sum_{i = 1}^{N} w_i h_i(x_s) \geq 0 \mbox{ for } s = 1, \ldots, S
 \end{equation}
 Thus, each training sample brings about an inequality, which demands to choose weights that are on one side of a "diagonal hyperplane" in $N$-dimensional space. The hyperplane is defined by a set of coefficients that are $\pm 1$ depending on the responses of the weak classifiers. Fig. ~\ref{fig:3d} illustrates the situation for $N=3$. The number of regions created by $S$ hyperplanes can be calculated using their characteristic polynomial \cite{Orlik}:
\begin{equation}
 N_{regions}= (-1)^N \sum_{S_{k}}(-1)^k (-1)^{dim(\bigcap S_{k})} \hbox{ , }
 \end{equation}
 where $S_{k}$ designates the $k$-element subsets of the $S$ hyperplanes and $dim(\bigcap S_{k})$ is the dimension of the intersection of $S_{k}$.\footnote{In this calculation we ignored the fact that a hyperplane or parts of it can become a solution space itself. This can occur when there are two training samples for which the sets of $\{ h_i(x_{s}) \}$ differ by a global sign but have the same label $y_{s}$. Since this case is exceedingly unlikely, the probability being of the order $O(S/{2^{(2N)}})$, we can afford not to consider this situation.} Due to linear dependencies among the hyperplanes, which occur for $N\geq 4$, we were not able to find a closed form expression for $dim(\bigcap S_{k})$ and instead have to resort to an upper bound for $N_{regions}$, which is known \cite{Orlik}\cite{Sauer72} to be
\begin{equation}
 N_{regions} \leq \sum_{k = 0}^{N} {S \choose k}
 \end{equation}
 It is possible that two different training samples generate identical inequality constraints for the $w_i$. In this sense, (7) is a conservative estimate as the actual number of solution spaces is often lower.
\section{Modifications to allow the application of the quantum adiabatic algorithm}
To bring (4) to a form that is amenable to AQC as implemented by the D-Wave hardware\footnote{The D-Wave hardware minimizes an Ising function via a physical annealing of thermal and quantum fluctuations.}, we need to effect several modifications. First, we need to transition from continuous weights $w_i \in [0, 1]$ to binary variables. Formally this can always be achieved by a binary expansion of the weights. The question that naturally arises is how many bits in the expansion are needed. Since each binary variable is associated with a qubit, it is important that we only use the minimal number necessary. Discrete weight configurations represented by a finite number of bits lie on a hypercubic lattice with edges that have $2^{bits}$ vertices. If each solution region contains a lattice vertex then all classifiers $H_{w}(x)$ that can be attained with real valued weights can also be realized by the discrete weight configurations. Thus one obtains a rough estimate for the required bit depth by demanding that the number of vertices on the $ (2^{bits})^N $ lattice is at least as large as the number of solution regions created by the hyperplanes $N_{regions}$. The weak classifiers are typically constructed in a way that only positive weights are needed. Hence, we only need to hit solution regions in the positive quadrant of which there are approximately $N_{regions}/2^N$.
\begin{eqnarray}
{ {Vertices \ on \ Lattice} \over {Regions \ in \ Positive \ Quadrant}} \approx {{(2^{bits})^N} \over {N_{regions}\over 2^N}}={{( 2^{bits+1})}^N \over N_{regions}} \\
\geq { { 2^{(bits+1)N} } \over {\sum_{k = 0}^{N}{S \choose k}}} \geq { { 2^{(bits+1)N} } \over {{({{e S}\over N})^N}} } = { { 2^{(bits+1)N} N^N} \over {{{(e S)}^N}} } {\buildrel {!}\over\geq}  1
\end{eqnarray}
\begin{eqnarray}
&\Rightarrow& {({{ 2^{(bits+1)} N} \over {e S}})}^N = {({{ 2^{(bits+1)} N} \over {e f N}})}^N = {({{ 2^{(bits+1)} } \over {e f }})}^N {\buildrel {!}\over\geq}  1 \\
&\Rightarrow& bits \geq {\log_{2} (f) + \log_{2} (e)-1} \mbox { , }
\end{eqnarray}
where $e$ is the Euler number and $f = {S \over N}$. On the left side of eqn. (9) we used a standard result regarding binomial coefficients: ${\sum_{k = 0}^{N}{S \choose k}} \leq {{({{e S}\over N})^N}}$. This holds in the case that $S \geq N$. Smaller numbers of training examples lead to an even lower bound than (11) for the required bit precision. Equation (11) is an important result as it shows that the bit precision needed for the weights only grows logarithmically with the ratio of the number of training examples to the number of weak classifiers. Thus for many problems that arise in practice we get away with very few bits and often we will only need only a single bit.

The second modification is not imposed by our desire to apply AQC per se but rather by the limitations of the D-Wave hardware, which calls for a Hamiltonian that has at most quadratic terms. To this end we effect a change in the loss function, now using the quadratic loss, such that finding $w^{opt}$ in (4) amounts to solving a quadratic optimization program:
\begin{eqnarray} w^{opt} = \arg\min_{w} \left(\sum_{s = 1}^{S} \vert \sum_{i = 1}^{N} w_i h_i(x_s) - y_s \vert^2 +
\lambda \parallel w \parallel _0 \right) \nonumber \\
= \arg\min_{w} \left(\sum_{s = 1}^{S} \left( \left( \sum_{i = 1}^{N} w_i h_i(x_s)\right)^2 -
2 \sum_{i = 1}^{N} w_i h_i(x_s) y_s + y_{s}^{2}\right) + \lambda \sum_{i = 1}^{N} {w_i}^0 \right)  \nonumber \\
= \arg\min_{w}
              \left(
                    \sum_{i = 1}^{N} \sum_{j = 1}^{N} w_i w_j
                    \underbrace{
		                \left(
                                      \sum_{s = 1}^{S} h_i(x_s) h_j(x_s)
                                 \right)
		               }_{Corr(h_i, h_j)}
                    +
                    \sum_{i = 1}^{N} w_i \left(
                                               \lambda - 2 \underbrace{
						                       \sum_{s = 1}^{S} h_i(x_s) y_s
                                                                      }_{Corr(h_i, y)}
                                         \right)
              \right)
\end{eqnarray}
In the third line we dropped $\sum_{s = 1}^{S} y_{s}^{2}$ because it represents a constant offset. In order for the square loss to be compatible with the binary decision enforced by the sign in eqn. (1) we scale the $h_i(x)$ such that $h_i:x \mapsto \{-{1\over N},{1\over N}\}$. Eqn. (12)
corresponds to a quadratic unconstrained binary optimization (QUBO) problem. Note that the transition from the second to the third line only holds for weights comprised of a single bit. If we use an arbitrary number of bits, we have to introduce an auxiliary bit $w_{i,aux}$ for each weight to enforce a 0-norm regularization within the framework of quadratic optimization. $R(w)$ of (4) then becomes
\begin{equation}
 R(w)=\sum_{i = 1}^{N} \kappa w_{i} (1-w_{i,aux}) + \lambda w_{i,aux}
\end{equation}
Minimizing $R(w)$ causes the $w_{i,aux}$ to act as indicator bits that are 1 when $w_{i}>0$ and 0 otherwise. For this to work $\kappa$ has to be chosen sufficiently large.

There is an intuitive way to look at (12). The weak classifiers $h_i$,
whose output is well correlated with the labels $y$ cause the bias term to be lowered, thus causing an increase
in the probability that $w_i = 1$. The coupling terms are proportional to the correlation among the weak classifiers.
Weak classifiers that are strongly correlated with each other cause the coupling energy to go up, thereby increasing
the probability for one of the correlated classifiers to be switched off, i.e. that either $w_i$ or $w_j$
becomes 0.
\par The matrix $Corr(h_i, h_j)$ figuring in the quadratic term is positive semi-definite,
thereby seemingly making the resulting optimization problem efficiently solvable with classical optimization techniques.
However, it has been confirmed that the quadratic unconstrained program with binary weights -- an integer programming problem -- is NP-hard, which
validates the motivation for applying quantum algorithms to find $w^{opt}$ in the above formulation \cite{Katayama}\cite{Helmberg}. Moreover, the matrix figuring in the quadratic term ceases to be of Gram type when the $w_i$ are
represented by more that one bit and the modified regularization term (13) leads to additional entries in the matrix.

\section{Implementation details}
\par We implemented the training formulations given by (4) and (12) in Matlab.
The dictionaries of weak classifiers that we employed consist of decision stumps of the form:
\begin{eqnarray} h_l^{1+}(x) &=& sign(x_l - \Theta_l^+)  \mbox{ for }  l = 1, \ldots, M \\
h_l^{1-}(x) &=& sign(-x_l - \Theta_l^-)  \mbox{ for }  l = 1, \ldots, M \\
h_l^{2+}(x) &=& sign(x_i x_j - \Theta_{i,j}^+)  \mbox{ for }   l = 1, \ldots, {M \choose 2}; i, j = 1, \ldots, M; i < j \\
h_l^{2-}(x) &=& sign(-x_i x_j - \Theta_{i,j}^-) \mbox{ for }   l = 1, \ldots, {M \choose 2}; i, j = 1, \ldots, M; i < j
\end{eqnarray}
Here $h_l^{1+}$, $h_l^{1-}$, $h_l^{2+}$ and $h_l^{2-}$ are positive and negative
weak classifiers of orders 1 and 2; $M$ is the dimensionality of the input vector $x$; $x_{l}$,$x_{i}$,$x_{j}$ are the elements of the input vector and $\Theta_l^+$, $\Theta_l^-$, $\Theta_{i,j}^+$ and
$\Theta_{i,j}^-$ are optimal thresholds of the positive and negative weak classifiers of orders 1, and 2 respectively. The input vectors are normalized using the 2-norm, i.e. we have
$\parallel x \parallel _2=1$.

\par Using the training data,
an approximately optimal threshold $\Theta$ is computed for each classifier. The goal is to obtain an operating point that
results in the minimum number of errors due to that weak classifier alone when the weak classifier is evaluated on the
entire training set.

\par To minimize (4) for the purpose of determining the optimal weights, we employ simulated annealing. An exponential cooling schedule
is used, and the schedule is tuned to the dataset for improved performance.

\par The QUBO from (12), can be rewritten as $w^{opt} = \arg\min_{w} \left( \sum_{i,j} Q_{i,j} w_i w_j \right) $, where
the coefficient matrix has elements $Q_{i,j} = Corr(h_i,h_j)$ and $Q_{i,i} = {S\over N^2} + \lambda - 2Corr(h_i,y)$. The resultant problem is solved
with a multi-start tabu solver tuned to QUBO problems \cite{Palubeckis2004}.

We noticed that we could achieve enhanced results by adding a post-processing step. The $w^{opt}$ returned by tabu search is used to compute an optimal threshold for the final strong classifier:
\hbox{$T=\frac{1}{S}\sum_{s=1}^S \sum_{i=1}^{N} w_i^{opt}h_i(x_s)$}, where the $x$ vectors are taken from a validation data set.
$T$ represents the average of all computed responses of the strong classifier immediately before the categorical decision is made. We modify eqn. (1) by inserting $T$. Thus the final classifier becomes
\begin{equation} y = {\rm sign}\displaystyle \left(\sum_{i = 1}^{N} w_i^{opt} h_i(x) - T \right) \end{equation}

After that, the set of test examples is evaluated using the strong classifier configured in this way
and the test errors are counted. The output consists of the
number of test errors, the number of weak classifiers with non-zero weights that make up the final strong classifier. Since the optimal regularization strength cannot be known a-priori for different data,
we use 30-fold cross validation in order to find a regularization strength $\lambda$ that results in the best generalization on a validation set. We only consider values of $\lambda$ for which the total number of weak classifiers does not exceed $N/2$.

\section{Performance measurements on benchmark problems}
\par
\begin{figure} [t]
\begin{center}
      \subfloat{\includegraphics[scale=0.4]{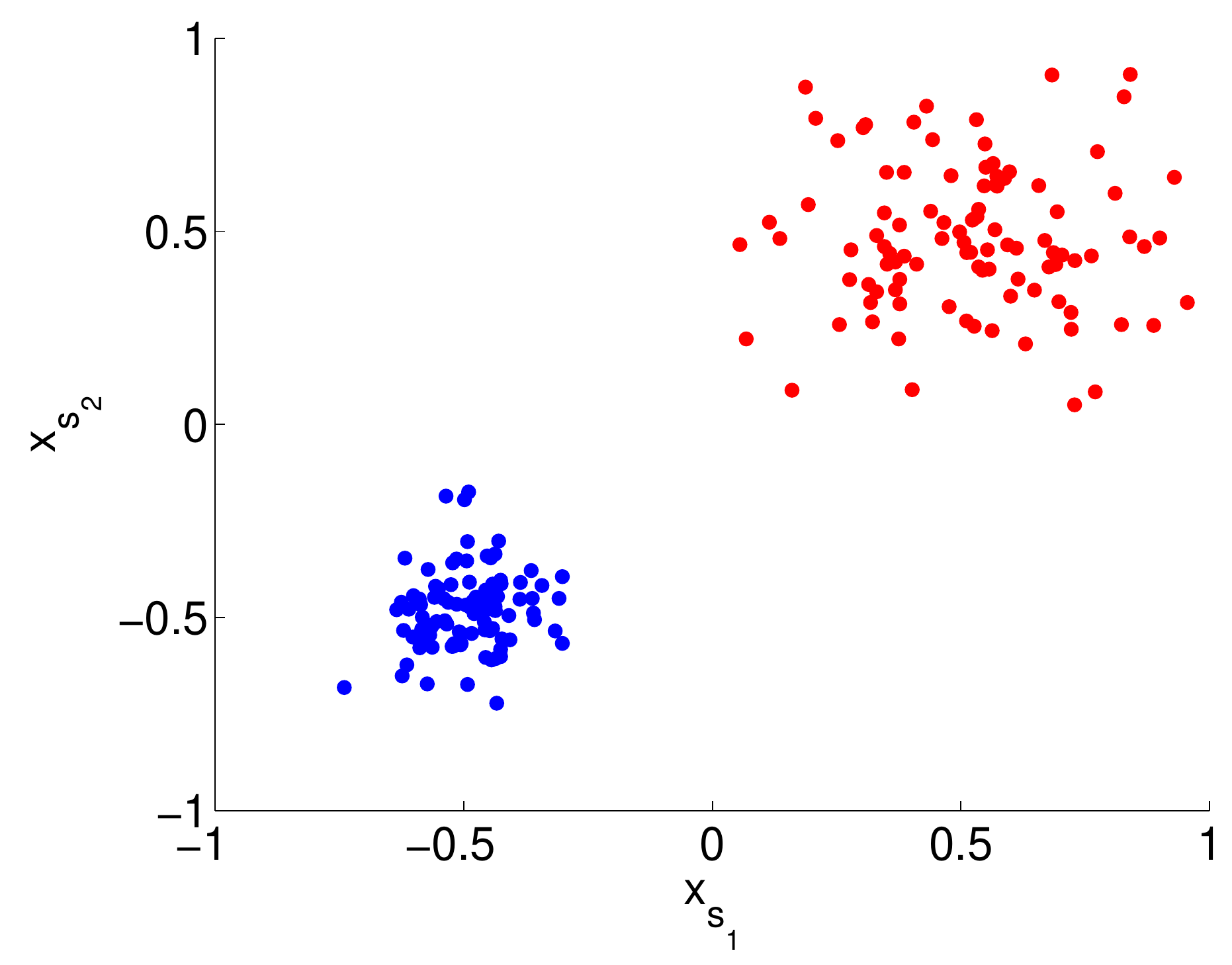}}
      \subfloat{\includegraphics[scale=0.4]{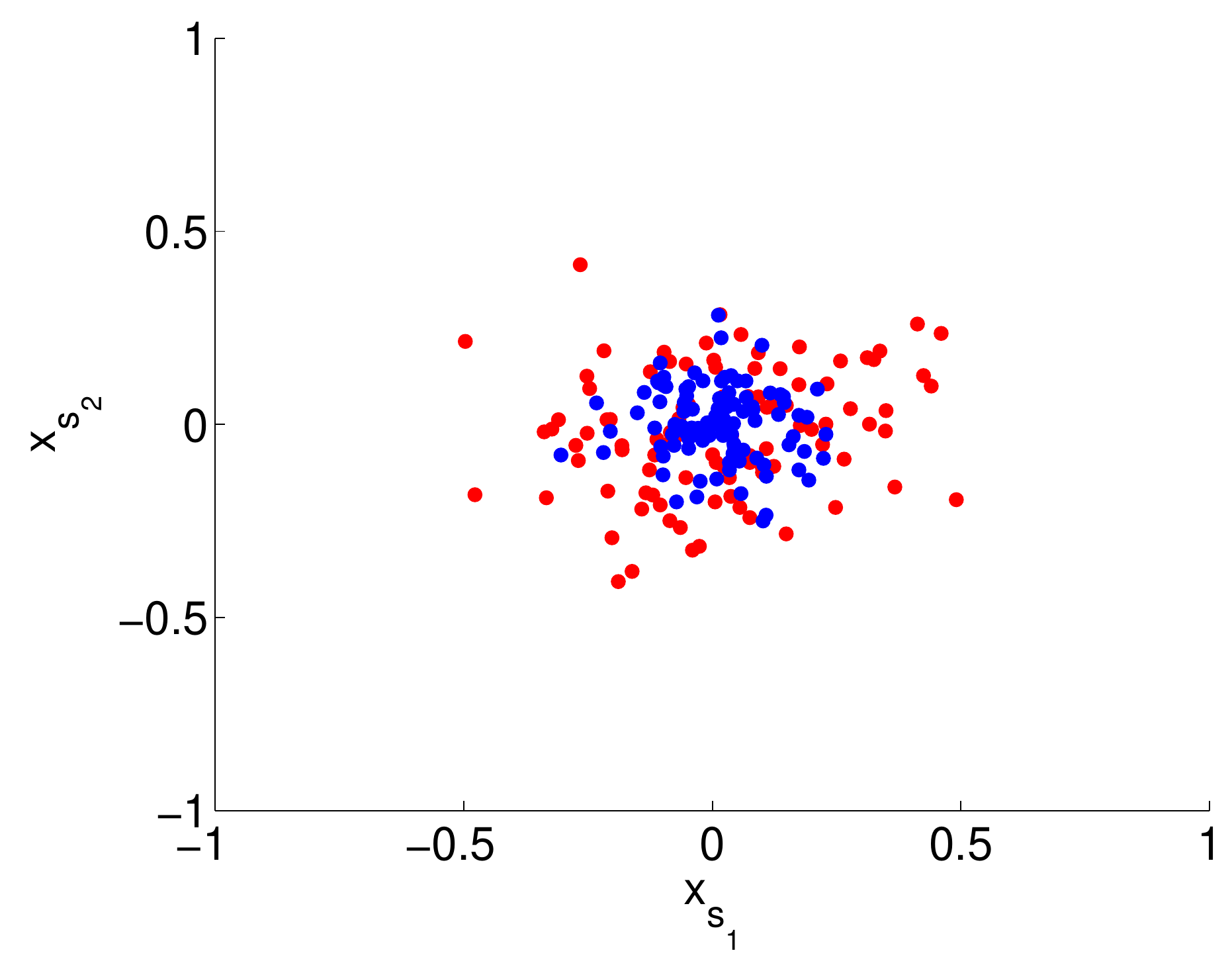}}
      \caption{Visualization of the synthetic test data employed in the benchmark tests. The data consists of
        two isotropic Gaussian distributions of 30-dimensional inputs with different variances. Shown are the first two dimensions.
        The two clouds represent the positive and negative data sets which are translated relative to each other.
        The position of their means is controlled by an overlap parameter chosen such that the clouds are maximally segregated
        when this overlap is 0 (left) and maximally overlap when the overlap is 1 (right). Thus, the classification task is hardest when the overlap parameter is 1.}
      \label{fig:synth}
\end{center}
\end{figure}

 To assess the performance of binary classifiers of the form (18) trained by solving the
optimization problems (4) and (12) respectively, we measured their test errors on 30-dimensional
synthetic and natural data sets. Synthetic test data was generated by sampling from $P(x,y) = {1\over2} \delta(y-1) N(x|\mu_+,I) + {1\over2} \delta(y+1) N(x|\mu_-,I)$ where $N(x|\mu,\Sigma)$ is a spherical Gaussian having mean $\mu$ and covariance $\Sigma$. An overlap coefficient determines the separation of the two Gaussians. The synthetic data is illustrated on Fig. ~\ref{fig:synth}. The natural data consists of vectors of Gabor wavelets amplitudes extracted at eye locations in images showing faces.
\par For comparison we ran the same tests using different implementations of boosting. First we implemented
AdaBoost as formulated by Freund and Shapire in \cite{Freund}. Here we used the same weak classifiers as
the ones defined in (14)-(17). Additionally, we compared the performance against the GML AdaBoost Matlab
Toolbox \cite{Vezhnevets}. The GML toolbox contains implementations of three different flavors of AdaBoost.
We ran the tests on all three of them but only display the test results for the best out of the three. The
GML toolbox is limited to the equivalent of our order 1 weak classifiers---classification and regression trees
with branching factor of 1. In the tests we varied the dictionary of weak classifiers, the bit precision used
to represent the weights as well as $f$, the ratio of training samples to weak classifiers. We used two
dictionaries. The first, called the order 1 dictionary, consists of the set of decision stumps with linear
arguments $h_l^{1+}$, $h_l^{1-}$ as per eqns. (14) and (15). The second one, called the order 2 dictionary, uses
the set of weak classifiers of the order 1 dictionary but adds the order 2 decisions stumps $h_l^{2+}$ and
$h_l^{2-}$ described in eqns. (17) and (18) as well. The order 1 dictionary has 60 weak classifiers while the
order 2 dictionary employs 930.
\begin{figure}[t!]
\begin{center}
      \subfloat{\includegraphics[scale=0.4]{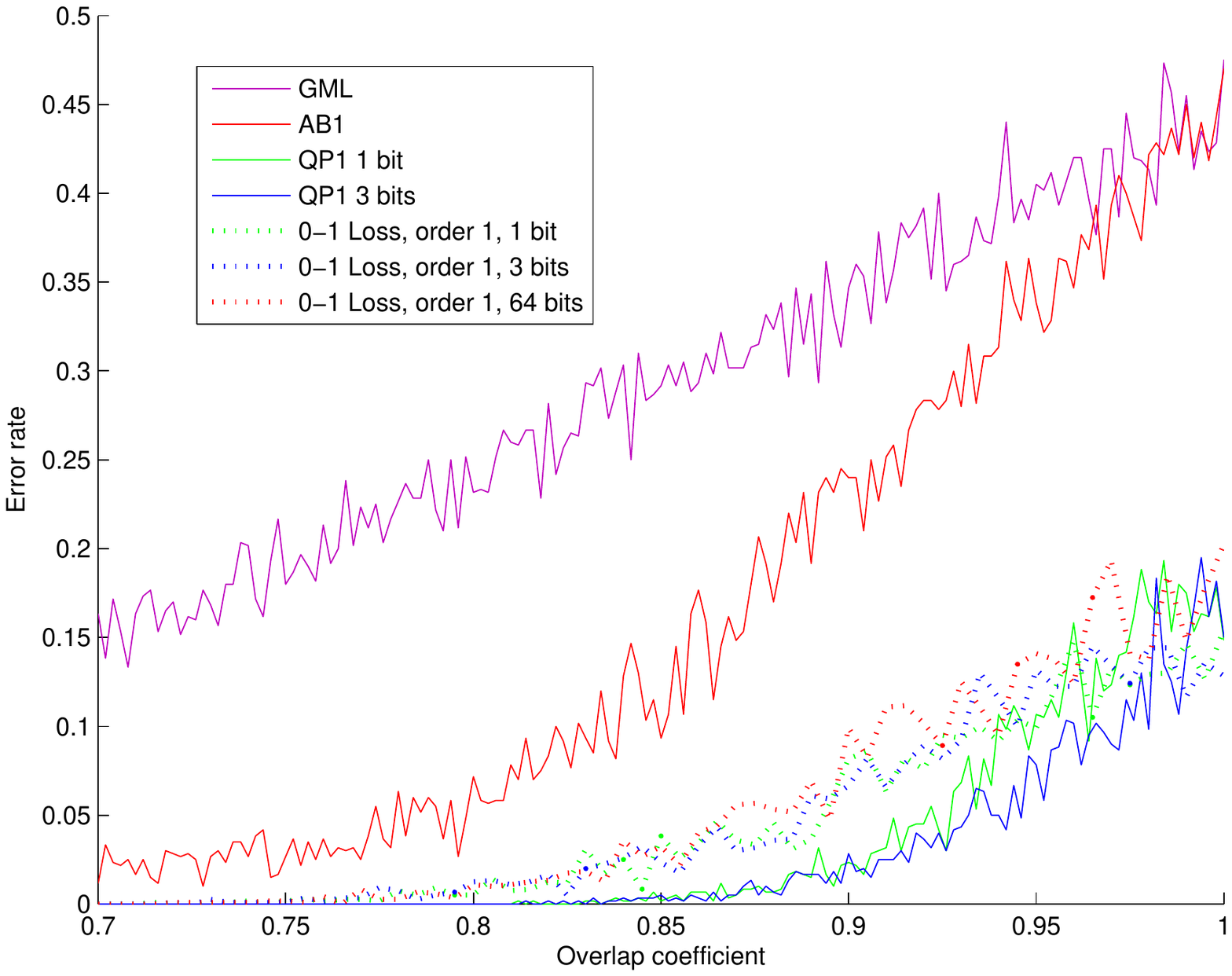}}
      \subfloat{\includegraphics[scale=0.4]{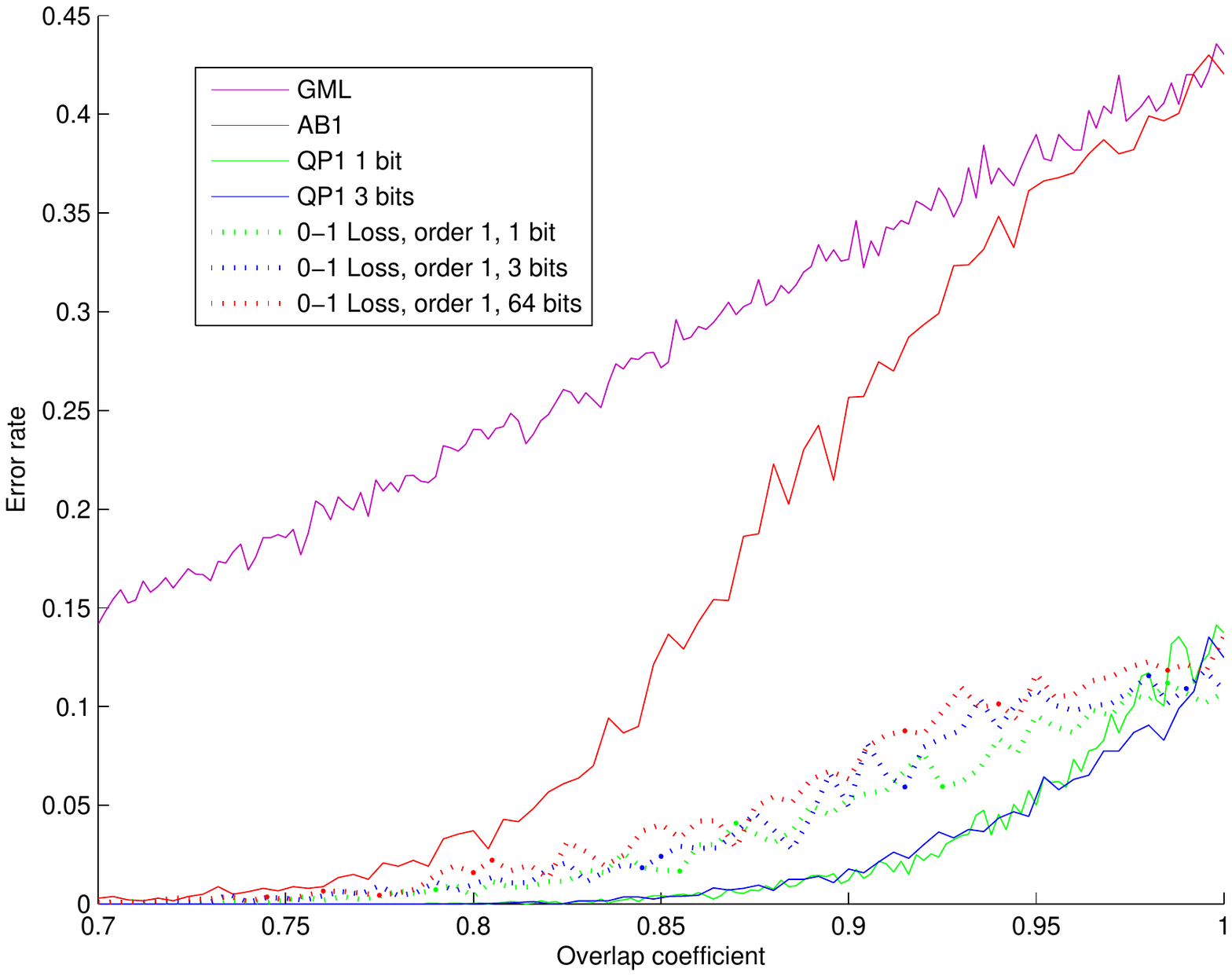}}
      \caption{Test errors for the synthetic data set with $f=1$ and $f=8$ for the {order 1} dictionary. Note that due to the different variances of the Gaussian distributions for the positive and negative training samples the generalization error does not necessarily approach 0.5 as the overlap becomes maximal.}
      \label{fig:synth1}
\end{center}
\begin{center}
      \subfloat{\includegraphics[scale=0.4]{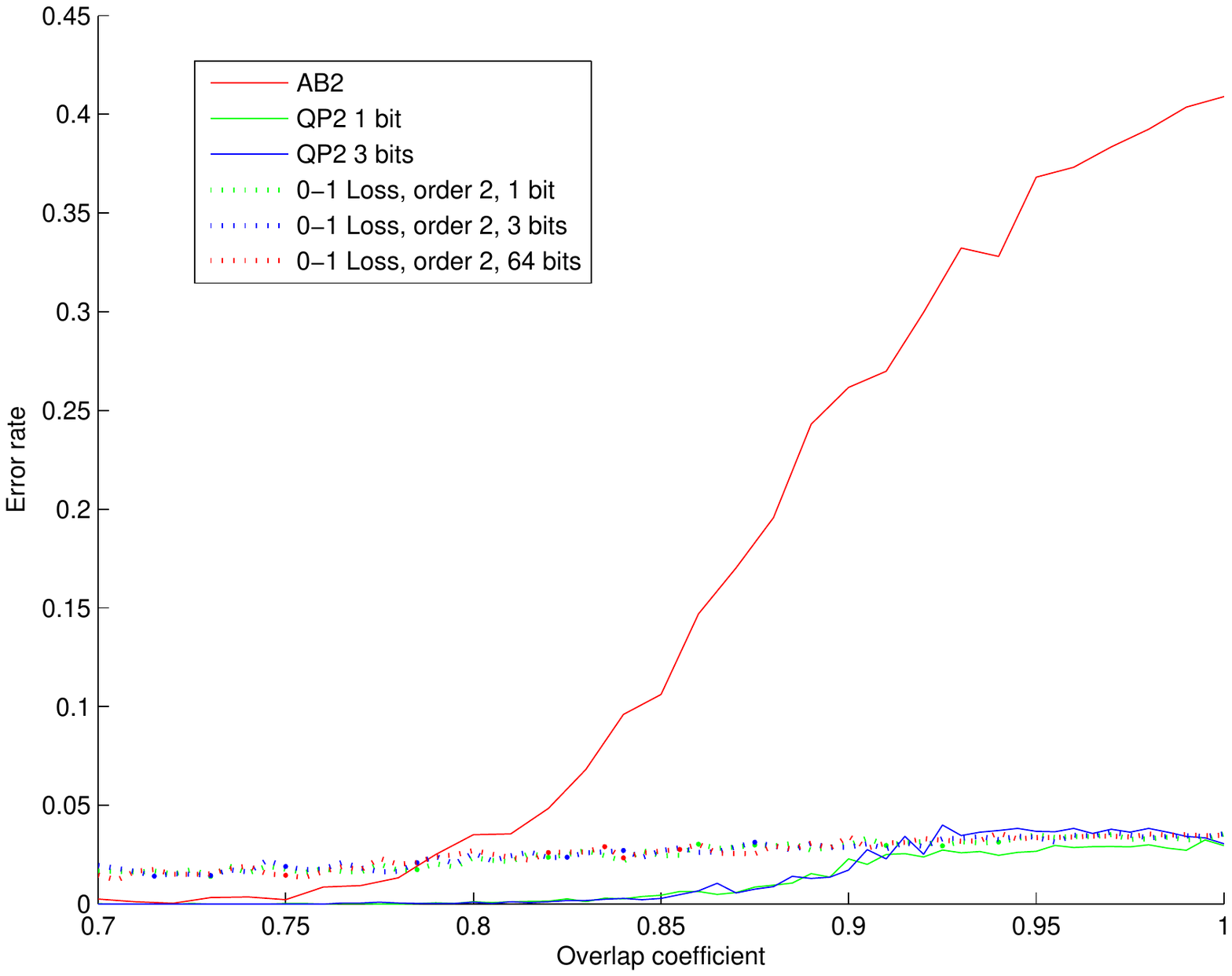}}
      \subfloat{\includegraphics[scale=0.4]{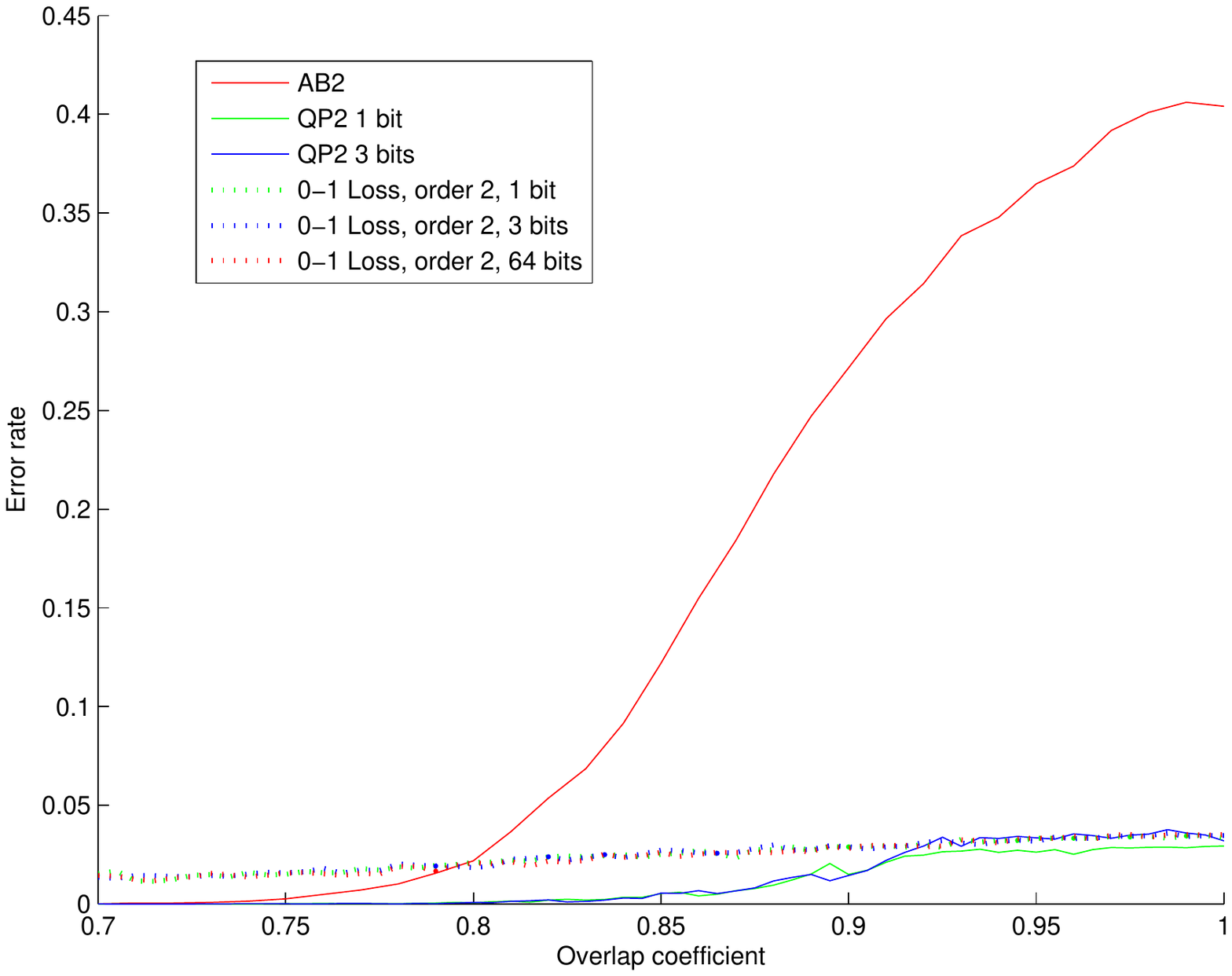}}
      \caption{Test errors for the synthetic data set with $f=1$ and $f=8$ for the {order 2} dictionary}
      \label{fig:synth2}
\end{center}
\end{figure}
\par Figs. ~\ref{fig:synth1} and ~\ref{fig:synth2} show the test errors on the synthetic data set we obtained for different configurations. QP1 and QP2 denote the classifiers trained with the quadratic program (12) while using the dictionaries of order 1 and 2, respectively. Similarly, 0-1 Loss 1 and 0-1 Loss 2 stand for the classifiers using dictionaries 1 and 2 trained by solving the optimization problem (4). AB1 and AB2 denote classifiers trained for the same dictionaries but with the AdaBoost algorithm. Finally, GML represents the best result obtained with the GML AdaBoost Matlab toolbox. For GML, only a dictionary equivalent to our order 1 dictionary is available. The figures show test errors obtained on data sets that were not used during the training but were drawn however i.i.d. from the the same distributions. Test error is plotted against overlap coefficient for the range 0.7 to 1 corresponding to an increasingly harder classification problem. Accordingly, we observe an increasing error rate. A number of observations can be made. First, the classifiers trained with global optimization outperform those obtained by the greedy feature selection methods employed in boosting. Second, with the exception of QP1, the global optimizations that used fewer bits to represent the weights did better than those that employed more. Classifiers using the richer order 2 dictionary achieved a lower test error, which, given the structure and size of the dictionaries relative to the input dimensionality is not surprising. We do not draw conclusions from the fact that 0-1 loss fared worse than quadratic loss. This could be an artifact of using simulated annealing to solve (4), while tabu search was employed to optimize (12).

\begin{figure} [b!]
\begin{center}
      \includegraphics[scale=0.75]{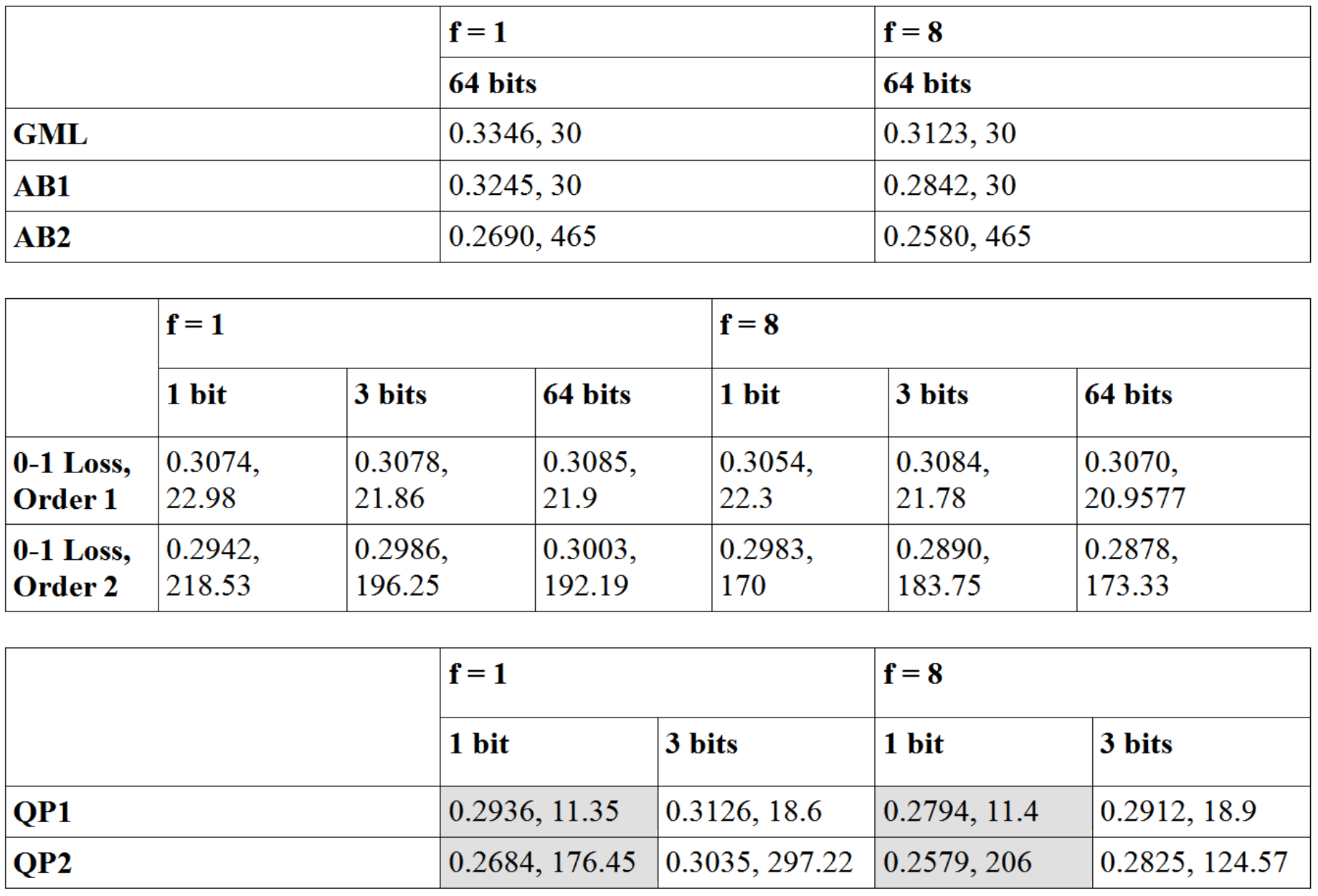}
      \caption{Test results obtained for a natural data set, which consisted of vectors of Gabor wavelets amplitudes
        extracted at eye locations in images showing faces. Each data cell in the tables contains two numbers -- the
        first represents the respective error rate, and the second gives the number of weak classifiers with a non-zero
        weight. The values are averages obtained through cross-validation of 100 runs. The shaded cells indicate the
        most accurate results.}
      \label{fig:natural}
\end{center}
\end{figure}

\par Besides synthetic data, we also did testing with natural data. The table in Fig. ~\ref{fig:natural}
shows the results obtained from a test set consisting of vectors of Gabor wavelets amplitudes
extracted at eye locations in images showing faces. The data consisted of 20,000 input vectors, which
we divided evenly into a training set, a validation set to fix the parameter $\lambda$, and a test
set. For QP the bit-constrained learners always performed better in terms of accuracy and classifier compactness. In the case of 0-1 Loss the performance was similar for the different bit depths with small trade-offs between accuracy and compactness. The global optimization approach using the quadratic objective function (12) yields the best results.
The accuracy is only increased by less than $10\%$ relative to AdaBoost, but this is accomplished with
a reduction of more than $50\%$ of the switched-on weak classifiers.

\section{Discussion}
\par We have seen an impressive performance of global optimization
approaches that minimize a regularized measure of training
error to find an optimal combination of weights for constructing a
binary classifier. Global optimization competes successfully with greedy methods such as
the state-of-the-art method AdaBoost. Further, we discovered that
bit-constrained learning machines often exhibit a generalization error that
is lower than the one obtained when the weights are represented with higher
precision. To the best of our knowledge, this has not been studied before.
Bit constraining can be regarded as an intrinsic regularization that
contributes to keeping the model complexity low. The finding that the
bit-precision needed to realize the optimal training error only grows
logarithmically with the ratio of the number of training examples to weak
classifiers, supplies insight into why few-bit learning machines work.
The competitive performance of bit-constrained classifiers suggests that
training benefits from being treated as an integer program. This has a
twofold implication. First, this is good news for hardware-constrained
implementations such as cell phones, sensor networks, or early quantum chips
with small numbers of qubits. Second, this renders the training problem
manifestly NP-hard, thus further motivating the application of quantum
algorithms that may generate better approximate solutions than classically
available. Our next steps will be to investigate the advantages that global
optimization with AQC hardware offers for our problem instances. We plan to
use the next generation of D-Wave chips with 128 qubits. This will involve
adjusting our implementation to additional engineering constraints of the
existing AQC hardware such as a sparse connectivity graph among the qubits.
Employing AQC during the training phase has the significant benefit that
once the optimal set of weights has been computed, then those can be taken
advantage of by an entirely classical processor. In this work we only
considered fixed dictionaries of weak classifiers. An important
generalization that remains to be studied is to apply this framework to
adaptive dictionaries. We want to conclude with the remark that our finding
that bit-constraint learning has good generalization
properties may have implications when studying plasticity in the nervous
system, where it is still an unresolved problem how a synapse can store
information reliably over a long period of time\cite{Kandel}.

\section*{Acknowledgments}
We would like to thank Hartwig Adam, Jiayong Zhang and Xiaowei Li for their help with preparing the natural test data; Alessandro Bissacco for his assistance with Matlab and reviewing the boosting code; Edward Farhi, Yoram Singer, Ulrich Buddemeier and Vint Cerf for commenting on earlier versions of the paper.

\bibliography{references}
\bibliographystyle{alpha}
\end{document}